# Density of states near the Anderson transition in a space of dimensionality $d=4-\epsilon$

I. M. Suslov

*P.L. Kapitsa Institute of Physical Problems, Russian Academy of Sciences, 117334 Moscow, Russia*

Asymptotically accurate results have been obtained for the average Green's function and the density of states in a Gaussian random potential for dimensionality of space $d=4-\epsilon$ over the entire energy region, including the vicinity of the mobility threshold. For $N\sim 1$ ($N$ is the order of the perturbation theory) only parquet terms corresponding to higher terms in $1/\epsilon$ are taken into account. For large $N$ all powers of $1/\epsilon$ are taken into account with their coefficients calculated in the main asymptotic limit in $N$. This calculation is performed by combining the condition of renormalization theory with the Lipatov asymptotic limit. © *1997 American Institute of Physics.* [S1063-7761(97)02605-X]

## 1. INTRODUCTION

According to generally accepted thinking,[1,2] the single-electron density of states does not have a singularity at the Anderson transition, in contrast to the conductivity and the localization radius of the wave functions.[3–6] Nevertheless, its calculation is of fundamental significance since all known methods break down in the vicinity of the transition. In addition, the density of states and the conductivity, defined respectively by the average Green's function $\langle G(x,x')\rangle$ and the correlator $\langle G^R G^A\rangle$, are not completely independent. A study in the parquet approximation shows[7] that the mathematical difficulties in both cases are of the same nature and are connected with the "ghost" pole problem. On the other hand, to satisfy the Ward identity linking the eigenenergy part with the irreducible vertex in the Bethe–Salpeter equation[8] would require exact agreement of the diagrams taken into account in the calculation of the conductivity and density of states; this circumstance is not dealt with in any of the presently existing theories[7] with the exception of the theory recently proposed in Ref. 9.

For weak disorder the mobility threshold lies in the vicinity of the starting boundary of the spectrum, at which the random potential can be taken to be Gaussian by virtue of the possibility of averaging over scales that are small in comparison with the wavelength of the electron, but large in comparison with the distance between scatterers (the so-called Gaussian segment of the spectrum[10]). Calculation of the average Green's function for the Schrödinger equation with Gaussian random potential reduces to the problem of a second-order phase transition with an $n$-component order parameter $\varphi=(\varphi_1,\varphi_2,\ldots,\varphi_n)$ in the limit $n\to 0$.[11,12] In this case the coefficients in the Ginzburg–Landau Hamiltonian

$$H\{\varphi\}=\int d^d x\left(\frac{1}{2}c|\nabla\varphi|^2+\frac{1}{2}\kappa_0^2|\varphi|^2+\frac{1}{4}u|\varphi|^4\right) \quad (1)$$

are linked with the parameters of the disordered system by the relations

$$c_0=1/2m,\quad \kappa_0^2=-E,\quad u=-a_0^d W^2/2, \quad (2)$$

where $d$ is the dimensionality of the space, $m$ and $E$ are the mass and energy of the particle, $a_0$ is the lattice constant, and $W$ is the amplitude of the random potential (in what follows we take $c_0=1$). The "incorrect" sign of the coefficient of $|\varphi|^4$ leads to the inapplicability of the usual mean-field theory and the necessity of a fluctuational treatment[11,13] over the entire parameter space; the functional integrals for $u<0$ are understood in the sense of an analytic continuation from positive $u$, which for a retarding Green's function is carried out through the lower half-plane.[12]

The present paper completes the program of constructing a $(4-\epsilon)$-expansion initiated in Refs. 14–16. The dimensionality of the space $d=4$ is singled out for the Hamiltonian (1) from considerations of renormalizability: for $d>4$ the theory is not renormalizable and the discreteness of the lattice is of fundamental significance, ensuring the existence of a cutoff parameter $\Lambda\sim a_0^{-1}$ at high momenta[14]; for $d=4$ a logarithmic situation holds sway, admitting the existence of both non-renormalizable[15] and renormalizable models[16]; for $d<4$ the theory is renormalizable with the help of one subtraction, and passage to the continuum limit $a_0\to 0$, $a_0^d W^2\to$const is possible. The use of simplifications arising at high dimensionalities to construct a $(4-\epsilon)$-dimensional theory requires the successive consideration of all four types of theories; this was done in Refs. 14–16 and in the present work. The results of this work have already been published in a brief exposition in Ref. 17.

## 2. STRUCTURE OF THE APPROXIMATION

The calculation of the average Green's function $\langle G(p,\kappa)\rangle$ ($p$ is the momentum and $\kappa$ is the renormalized value of $\kappa_0$) reduces in the standard way to a calculation of the eigenenergy $\Sigma(p,\kappa)$, for which the structure of the perturbation-theory series in four-dimensional space at $p=0$ has the form[15]

$$\Sigma(0,\kappa)-\Sigma(0,0)=\kappa^2\sum_{N=1}^{\infty}u^N\sum_{K=0}^{N}A_N^K\left(\ln\frac{\Lambda}{\kappa}\right)^K. \quad (3)$$

Reference 16 established the structure of the approximation which allows one to obtain asymptotically accurate results (in the limit of weak disorder) for a renormalizable class of models, this being the zeroth approximation for the $(4-\epsilon)$-dimensional theory. For $N\sim 1$ it is sufficient to take account of the coefficients $A_N^N$ corresponding to the "leading logarithms;" for large $N$ this approximation is insufficient in light of the higher rate of growth with respect to $N$ of the coefficients of the lower-order logarithms: therefore it is generally speaking necessary to take account of all the coefficients $A_N^K$, but it suffices to calculate them in the leading asymptotic limit in $N$. The latter is possible by combining the condition of renormalizability of the theory with the Lipatov asymptotic limit.[18]

The sum of the high-order terms of the perturbation-theory series gives a nontrivial contribution associated with the divergence of the series and is important only for negative $u$; this latter result explains why in the usual theory of phase transitions it is possible to restrict the calculation to the leading logarithmic approximation.[19,20]

For $d=4-\epsilon$ the expansion analogous to (3) has the form

$$\kappa^2 + \Sigma(0,\kappa) - \Sigma(0,0) \equiv \kappa^2 Y(\kappa)$$

$$= \kappa^2 \sum_{N=0}^{\infty} (u\Lambda^{-\epsilon})^N \sum_{K=0}^{N} A_N^K(\epsilon)$$

$$\times \left[ \frac{(\Lambda/\kappa)^\epsilon - 1}{\epsilon} \right]^K, \quad (4)$$

where the coefficients $A_N^K(\epsilon)$ are finite in the limit $\epsilon \to 0$ and $A_0^0(\epsilon) \equiv 1$. Expansion (4) follows from the fact that the quantity $Y$ in $N$th-order perturbation theory is a homogeneous polynomial of degree $N$ built up from $\Lambda^{-\epsilon}$ and $\kappa^{-\epsilon}$: indeed, in the transition from the $N$th-order diagram to the $(N+1)$-th–order diagram the dimensionality in the momentum decreases by $\epsilon$ (Ref. 21), which gives the factor $\Lambda^{-\epsilon}$ or $\kappa^{-\epsilon}$ depending on whether high or low momenta determine the corresponding contribution. Separating out the factor $\epsilon^{-K}$ ensures the correct limit in expansion (3) as $\epsilon \to 0$.

The standard procedure for carrying out the $\epsilon$-expansion[11,13] consists in expanding the coefficients $A_N^K(\epsilon)$ in powers of $\epsilon$

$$A_N^K(\epsilon) = \sum_{L=0}^{\infty} A_N^{K,L} \epsilon^L \quad (5)$$

and preserving in each order of the perturbation theory some of the higher orders in $1/\epsilon$; the first $\epsilon$-approximation corresponds to taking account of only the coefficients $A_N^{N,0}$, which coincide with the coefficients of the leading logarithms in expansion (3). As is the case for $d=4$, such an approximation is insufficient for $u<0$ due to the higher rate of growth with $N$ of the coefficients of the lower terms in $1/\epsilon$: limiting the expansion to the coefficients $A_N^{N,0}$ is possible only for $N \sim 1$, whereas for larger $N$ it is necessary to take into account all the coefficients $A_N^{K,L}$, calculating them in the leading asymptotic limit in $N$.

According to Eq. (4) the quantity $Y$ is a function of $g_0 \equiv u\Lambda^{-\epsilon}$ and $\Lambda/\kappa$; it satisfies the Callan–Symanzik equation

$$\left( \frac{\partial}{\partial \ln \Lambda} + W(g_0,\epsilon) \frac{\partial}{\partial g_0} + V(g_0,\epsilon) \right) Y = 0, \quad (6)$$

which expresses the condition of renormalizability of the theory, and Eq. (15) of Ref. 16, which was obtained in an analogous way. The functions $W(g_0,\epsilon)$ and $V(g_0,\epsilon)$ can be expanded in the following series:

$$W(g_0,\epsilon) = \sum_{M=1}^{\infty} W_M(\epsilon) g_0^M = \sum_{M=1}^{\infty} \sum_{M'=0}^{\infty} W_{M,M'} g_0^M \epsilon^{M'},$$

$$V(g_0,\epsilon) = \sum_{M=1}^{\infty} V_M(\epsilon) g_0^M = \sum_{M=1}^{\infty} \sum_{M'=0}^{\infty} V_{M,M'} g_0^M \epsilon^{M'}, \quad (7)$$

whose first coefficients were calculated in Ref. 21:

$$W_1(\epsilon) = -\epsilon, \quad W_{2,0} = K_4(n+8),$$
$$W_{3,0} = -3K_4^2(3n+14), \quad V_{1,0} = -K_4(n+2) \quad (8)$$

(according to Ref. 16 the function $V(g_0,\epsilon)$ coincides with the function $\eta_2(g_0,\epsilon)$ introduced in Ref. 21); the quantity $K_4$ is defined in Eqs. (14). Substituting expansions (4) and (7) into Eq. (6) leads to a system of equations for the coefficients $A_N^K(\epsilon)$:

$$(K+1)A_N^{K+1}(\epsilon) = (N-K)\epsilon A_N^K(\epsilon)$$

$$- \sum_{M=1}^{N-K} [(N-M)W_{M+1}(\epsilon)$$

$$+ V_M(\epsilon)]A_{N-M}^K(\epsilon), \quad (9)$$

or for the coefficients $A_N^{K,L}$:

$$(K+1)A_N^{K+1,L} = (N-K)A_N^{K,L-1}(1-\delta_{L,0})$$

$$- \sum_{M=1}^{N-K} \sum_{M'=0}^{L} [(N-M)W_{M+1,M'}$$

$$+ V_{M,M'}]A_{N-M}^{K,L-M'}. \quad (10)$$

Wilson's method[11,13] is based on the fact that in the $n$th $\epsilon$-approximation one needs to know the coefficients $A_N^{N-K,L}$ for $K+L \leq n-1$, for which Eqs. (10) yield the closed system of difference equations

$$-Nx_N = [W_{2,0}(N-1) + V_{1,0}]x_{N-1},$$

$$-(N-1)y_N = [W_{2,0}(N-1) + V_{1,0}]y_{N-1} + [W_{3,0}(N-2)$$

$$+ V_{2,0}]x_{N-2},$$

$$-Nz_N = [W_{2,0}(N-1) + V_{1,0}]z_{N-1} + [W_{2,1}(N-1)$$

$$+ V_{1,1}]x_{N-1} - y_N, \quad (11)$$

(where $x_N \equiv A_N^{N,0}$, $y_N \equiv A_N^{N-1,0}$, $z_N \equiv A_N^{N,1}$, ...), which is solvable by the method of variation of parameters;[22] assigning the initial conditions and determining the quantities $W_{2,0}$, $V_{1,0}$, ... requires the calculation of some lower orders of the perturbation theory. In particular, for the coefficients $A_N^{N,0}$ we easily obtain

$$A_N^{N,0} = (-W_{2,0})^N \frac{\Gamma(N-\beta_0)}{\Gamma(N+1)\Gamma(-\beta_0)},$$

$$\beta_0 = -\frac{V_{1,0}}{W_{2,0}} = \frac{n+2}{n+8}. \quad (12)$$

To investigate the higher orders in $\epsilon$, the Wilson method turns out to be ineffective, and it is more convenient to start with (9). Information about the coefficients $A_N^K(\epsilon)$ for $N \gg 1$ can be obtained by the Lipatov method,[18] according to which the later coefficients of the expansion in $u$ of the functional integrals with the Hamiltonian (1) are determined by the saddle-point configurations—instantons—and have factorial growth in $N$. For factorial series there exists a simple algebra that enables one to manipulate them as simply as finite expressions,[15] which in turn enables one to find the expansion coefficients of arbitrary $M$-point Green's functions, proceed from them to the eigenenergy and the vertex parts, etc. According to Sec. 6, the $N$th coefficient of the expansion of $\Sigma(p,\kappa)$ in powers of $u$ has the form

$$[\Sigma(p,\kappa)]_N = c_2 \Gamma(N+b)$$

$$\times a^N \int_0^\infty d\ln R^2 R^{-2} \langle \phi_c^3 \rangle_{Rp} \langle \phi_c^3 \rangle_{-Rp}$$

$$\times \exp\left( -Nf(\kappa R) + N\epsilon \ln R \right.$$

$$\left. + 2K_d I_4(\kappa R) \frac{1-(\Lambda R)^{-\epsilon}}{\epsilon} \right), \quad (13)$$

where

$$a = -3K_4, \quad b = \frac{d+2}{2}, \quad c_2 = c(3K_4)^{7/2},$$

$$f(x) = -\frac{\epsilon}{2}(C+2+\ln\pi) - 3x^2\left(C + \frac{1}{2} + \ln\frac{x}{2}\right),$$

$$\langle \phi_c^3 \rangle_p = 8\sqrt{2}\pi^2 p K_1(p),$$

$$I_4(x) = \widetilde{I}_4 \exp(f(x)), \quad \widetilde{I}_4 = \frac{16}{3} S_4,$$

$$S_d = 2\pi^{d/2}/\Gamma(d/2), \quad K_d = S_d(2\pi)^{-d}, \quad (14)$$

$C$ is Euler's constant, $K_1(x)$ is the modified Bessel function, and the constant $c$ is defined below in Sec. VI. Re-expanding series (4)

$$\kappa^2 + \Sigma(0,\kappa) - \Sigma(0,0) = \kappa^2 \sum_{N=0}^\infty (u\kappa^{-\epsilon})^N \sum_{K=0}^N B_N^K(\epsilon)$$

$$\times \left[ \frac{1-(\Lambda/\kappa)^{-\epsilon}}{\epsilon} \right]^K, \quad (15)$$

in such a way that the coefficients $B_N^K(\epsilon)$ are related to the coefficients $A_N^K(\epsilon)$ by

$$A_N^K(\epsilon) = \sum_{K'=0}^K C_{N-K'}^{N-K} B_N^{K'}(\epsilon) \epsilon^{K-K'}, \quad (16)$$

setting $p=0$ in Eq. (13), making the substitution $R \to R/\kappa$, and transforming the exponential

$$\exp\left\{ 2K_d I_4(R) \frac{1-(\Lambda R/\kappa)^{-\epsilon}}{\epsilon} \right\}$$

$$= \exp\left\{ 2K_d I_4(R) \frac{1-R^{-\epsilon}}{\epsilon} \right\}$$

$$\times \sum_{K=0}^\infty \frac{\{2K_d I_4(R) R^{-\epsilon}\}^K}{K!} \left[ \frac{1-(\Lambda/\kappa)^{-\epsilon}}{\epsilon} \right]^K, \quad (17)$$

we obtain for the coefficients $B_N^K(\epsilon)$ at large $N$

$$B_N^K(\epsilon) = \widetilde{c}_2 \Gamma(N+b) a^N \frac{1}{K!} \int_0^\infty d\ln R^2 R^{-2}$$

$$\times (2K_d I_4(R) R^{-\epsilon})^K$$

$$\times \exp\left( -Nf(R) + N\epsilon \ln R + 2K_d I_4(R) \frac{1-R^{-\epsilon}}{\epsilon} \right), \quad (18)$$

where $\widetilde{c}_2 = c_2 \langle \phi_c^3 \rangle_0^2$. By analogy with the case $d=4$ (Ref. 16), the Lipatov method reproduces the coefficients $B_N^K(\epsilon)$ well only for $K \ll N$, which is connected with their rapid falloff with $K$ and the limited accuracy ($\sim 1/N$) of the leading asymptotic behavior. Substituting (18) into Eq. (16), we obtain the following result for the coefficients $A_N^K(\epsilon)$ with $N \gg 1$:

$$A_N^K(\epsilon) = \widetilde{c}_2 \Gamma(N+b) a^N C_N^K \int_0^\infty d\ln R^2 R^{-2}$$

$$\times \left( \epsilon + \frac{2K_d \widetilde{I}_4}{N} e^{f(R) - \epsilon \ln R} \right)^K$$

$$\times \exp\left( -Nf(R) + N\epsilon \ln R + 2K_d I_4(R) \frac{1-R^{-\epsilon}}{\epsilon} \right), \quad (19)$$

which follows from Eq. (18) under the condition that the sum in Eq. (16) is determined by values of $K' \ll N$. Retaining only the term with $M=1$ in the sum (9), it is easy to convince oneself that the equation so obtained is satisfied by the result (19) for $K \ll N$ in the case $N\epsilon \lesssim 1$ and for all $K$ for $N\epsilon \gg 1$. The latter has to do with the fact that for $N\epsilon \gg 1$, the sum in Eq. (16) is determined by values of $K' \sim K/\epsilon N \ll N$ for all $K$ in the region of applicability of formula (18). The indicated reduction of Eq. (9) is possible at large $N$ by virtue of the factorial growth of $A_N^K(\epsilon)$ under the assumption that $W_N(\epsilon)$ and $V_N(\epsilon)$ grow more slowly than $A_N^0(\epsilon)$. This latter result can be assumed to be a consequence of the validity of formula (19) for $K=0, 1, 2$ (see Ref. 16 for a more detailed exposition).

The system of equations (9) determines the coefficients $A_N^K(\epsilon)$ with $K>0$ for prescribed $A_N^0(\epsilon)$. Since Eq. (19) is valid for the latter for all $N \gg 1$, it can be used as a boundary condition on system of equations (9), which enables one to determine all the $A_N^K(\epsilon)$ with large $N$. Thus, retaining only the leading order in $1/\epsilon$ for $N \sim 1$, defined by the coefficients (12), it is not hard to find the sum of series (4).

### 3. STUDY OF THE COEFFICIENTS $A_N^K(\epsilon)$

We will limit the sum (9) to terms with $M=1$ and $M=2$:

$$KA_N^K(\epsilon) = (N-K+1)\epsilon A_N^K(\epsilon) - W_2(\epsilon)[N-1-\beta(\epsilon)]$$
$$\times A_{N-1}^{K-1}(\epsilon) - W_3(\epsilon) N A_{N-2}^{K-1}(\epsilon), \qquad (20)$$

Here

$$\beta(\epsilon) = -\frac{V_1(\epsilon)}{W_2(\epsilon)} \xrightarrow{\epsilon \to 0} \beta_0. \qquad (21)$$

We set $A_N^{N+1}(\epsilon) = 0$ by definition in order to account for the absence of the last term in Eq. (20) with $K=N$. The last term in Eq. (20) is of order $\sim 1/N$ in comparison with the previous term and is taken to lowest order in $1/N$; the need to take it into account has to do with the fact that to calculate $A_N^K(\epsilon)$ with $K \sim N$ from the assigned values of $A_N^0(\epsilon)$ requires $\sim N$ iterations, and for an accuracy of each iteration of $\sim 1/N$ the errors build up. In what follows we will drop the argument $\epsilon$ in the intermediate formulas.

Making the substitution

$$A_N^K = (-W_2)^K \frac{\Gamma(N-\beta)}{\Gamma(K+1)\Gamma(N-K-\beta)} A_{N-K}^0 X_{N,N-K} \qquad (22)$$

in Eq. (20) and introducing the notation

$$h_M = -\frac{\epsilon}{W_2} \frac{A_{M+1}^0}{A_M^0} \frac{M+1}{M-\beta},$$

$$f_M = \frac{W_3}{W_2} \frac{A_{M-1}^0}{A_M^0} (M-1-\beta), \qquad (23)$$

we obtain

$$X_{N,M} = h_M X_{N,M+1} + X_{N-1,M} + \frac{f_M}{N} X_{N-2,M-1} \qquad (24)$$

with boundary condition

$$X_{N,N} = 1. \qquad (25)$$

Rewriting Eq. (24) in the form

$$X_{N,M} = (\hat{l}_M + \hat{\delta}_M) X_{N,M+1}, \qquad (26)$$

where

$$\hat{l}_M \equiv h_M + e^{-i\hat{p}}, \quad \hat{\delta}_M \equiv \frac{f_M}{N} e^{-2i\hat{p}}, \qquad (27)$$

$e^{-i\hat{p}}$ is the shift operator by $-1$, which operates on both arguments, and invoking the boundary condition (25), it is easy to obtain

$$X_{N,M} = (\hat{l}_M + \hat{\delta}_M)(\hat{l}_{M+1} + \hat{\delta}_{M+1}) \ldots (\hat{l}_{N-1} + \hat{\delta}_{N-1}) X_{N,N}$$

$$= \hat{l}_M \hat{l}_{M+1} \ldots \hat{l}_{N-1} 1 + \sum_{p_1=M}^{N-1} \hat{l}_M \ldots \hat{l}_{p_1-1} \hat{\delta}_{p_1} \hat{l}_{p_1+1} \ldots \hat{l}_{N-1} 1$$

$$+ \sum_{p_1=M}^{N-2} \sum_{p_2=p_1+1}^{N-1} \hat{l}_M \ldots \hat{l}_{p_1-1} \hat{\delta}_{p_1} \hat{l}_{p_1+1} \ldots \hat{l}_{p_2-1}$$

$$\times \hat{\delta}_{p_2} \hat{l}_{p_2+1} \ldots \hat{l}_{N-1} 1 + \ldots . \qquad (28)$$

We do not indicate the argument $N$ of the operator $\hat{\delta}$, which is shown on the left side of the equation. The main contribution to the sum comes from terms with a small number of operators $\hat{\delta}$, which are not difficult to calculate. The result

$$\hat{l}_M \hat{l}_{M+1} \ldots \hat{l}_{M'-1} = \sum_{L=0}^{M'-M} C_{M'-M}^L h_M h_{M+1} \ldots h_{M'-L-1} e^{-iL\hat{p}}, \qquad (29)$$

which determines the zeroth-order term in $\hat{\delta}$, follows by induction. For products with one $\hat{\delta}$ operator we have

$$\hat{l}_M \ldots \hat{l}_{p_1-1} \hat{\delta}_{p_1} \hat{l}_{p_1+1} \ldots \hat{l}_{M'-1}$$

$$= \sum_{L_1=0}^{p_1-M} \sum_{L_2=0}^{M'-p_1-1} C_{p_1-M}^{L_1} C_{M'-p_1-1}^{L_2} h_M \ldots h_{p_1-L_1-1}$$

$$\times \frac{f_{p_1-L_1}}{N-L_1} h_{p_1-L_1-1} \ldots h_{M'-L_1-L_2-3} e^{-i(L_1+L_2+2)\hat{p}}. \qquad (30)$$

Noting that by virtue of (23)

$$f_{p_1-L_1} h_{p_1-L_1-1} = \left(-\frac{\epsilon W_3}{W_2^2}\right)(p_1-L_1), \qquad (31)$$

we reduce the result (30) to the form

$$\left(-\frac{\epsilon W_3}{W_2^2}\right) \sum_L h_M \ldots h_{M'-L-3} e^{-i(L+2)\hat{p}}$$

$$\times \sum_{L_1} C_{p_1-M}^{L_1} C_{M'-p_1-1}^{L-L_1} \frac{p_1-L_1}{N-L_1}. \qquad (32)$$

The sum over $L_1$ has a saddle-point at $L_c = L(p_1-M)/(M'-M-1)$; replacing $L_1$ by $L_c$ in the last fraction in (32) and making use of the addition theorem for binomial coefficients (Ref. 22, p. 745) we obtain

$$\left(-\frac{\epsilon W_3}{W_2^2}\right) \sum_L C_{M'-M-1}^L h_M \ldots h_{M'-L-3} e^{-i(L+2)\hat{p}}$$

$$\times \left.\frac{p_1-(p_1-M)\tau}{N-(p_1-M)\tau}\right|_{\tau=L/(M'-M-1)}. \qquad (33)$$

Result (33) has the same structure as (29), and by induction it is not hard to find products with a moderate number $s$ of $\hat{\delta}$ operators; from Eq. (28) we obtain

$$X_{N,M} = \sum_{s=0}^{\infty} \left(-\frac{\epsilon W_3}{W_2^2}\right)^s \sum_{L=0}^{\min\{N-M-s, N-2s\}}$$

$$\times C_{N-M-s}^L h_M \ldots h_{N-L-2s-1}$$

$$\times \sum_{p_1=M}^{N-s} \frac{p_1-(p_1-M)\tau}{N-(p_1-M)\tau} \sum_{p_2=p_1+1}^{N-s+1}$$

$$\times \frac{p_2-2-(p_2-M-1)\tau}{N-2-(p_2-M-1)\tau} \sum_{p_s=p_{s-1}+1}^{N-1}$$

$$\times \left.\frac{p_s-2s+2-(p_s-M-s+1)\tau}{N-2s+2-(p_s-M-s+1)\tau}\right|_{\tau=L/(N-M-s)}. \qquad (34)$$

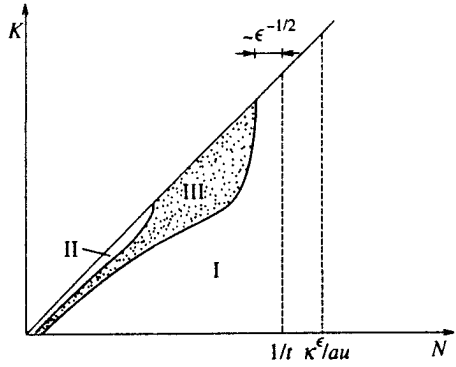

FIG. 1. Regions I and II, which give nonperturbative and quasiparquet contributions to the sum (4); the parameter $t \sim \epsilon$ is defined by formula (42). The nonperturbative contribution is estimated in effect for $N = \kappa^\epsilon/au$; the inequality $\kappa^\epsilon/au > 1/t$ corresponds to a positive value of $\Delta$ (see Eq. (45)).

Calculating the sum over $p_1, p_2, \ldots, p_s$ in the two overlapping regions of parameter space, we obtain the following results for it ($\tau' \equiv 1 - \tau$):

$$\frac{1}{s!}\left\{\frac{M+\tau'(N-M)}{(1-\tau')^2}\ln\frac{N}{\tau'N+(1-\tau')M}\right.$$

$$\left.-\frac{\tau'(N-M)}{1-\tau'}\right\}^s, \quad \max\{\tau'N,M\}\gg s, \qquad (35)$$

$$\frac{1}{s!}\frac{\Gamma(M+\tau'N+1)}{\Gamma(M+\tau'N-s+1)}\left\{\ln\frac{N}{\tau'N+M-s}\right.$$

$$\left.-\frac{\tau'N}{\tau'N+M}\right\}^s, \quad \tau'N \sim M \sim s. \qquad (36)$$

In the first case it is possible to neglect quantities $\sim s$ in the fractions within the summation range, and transform from sums to integrals; in the second case it is possible to calculate the sums systematically by separating out the two highest powers of the large logarithms. Formula (36) is valid literally for $s \gg 1$, whereas for $s \sim 1$ the difference between the expression in braces and $\ln N$ exceeds the accuracy of the calculation.

The product

$$h_M h_{M+1} \ldots h_{N-L-2s-1} = \left(-\frac{\epsilon}{W_2}\right)^{N-L-2s-M}\frac{A^0_{N-L-2s}}{A^0_M}$$

$$\times\frac{\Gamma(M-\beta)}{\Gamma(M+1)}\frac{\Gamma(N-L-2s+1)}{\Gamma(N-L-2s-\beta)} \qquad (37)$$

entering into expression (34) depends on the coefficients $A^0_N$, which are assumed to be known. By analogy with the case $d = 4$ (Ref. 16), in the $(N,K)$ plane it is possible to distinguish two regions in Fig. 1: region I, in which the sum in (34) is determined by indices $N - L - 2s \gg 1$, such that the Lipatov asymptotic limit is valid for the coefficients $A^0_N$, and region II ($M \ll \ln N$, $N\epsilon \ll 1$), "controlled" by the trivial coefficient $A^0_0 = 1$. Between regions I and II lies the region of non-universality—region III ($M \sim \ln N$, $N\epsilon \lesssim 1$), in which information about the coefficients $A^0_N$ with $N \sim 1$ is important. Region III does not make a substantial contribution to the sum (4).

The conditions $N - L - 2s \gg 1$, $\max\{\tau'N,M\} \gg 1$, and $N - M - L \gg s$ are satisfied in region I for $N\epsilon \gtrsim 1$. This enables one to use (19) for $A^0_N$ and (35) for the sum over $p_i$, and neglect the magnitude of $s$ in the slowly varying functions within the summation range in (34) and to sum over $s$. After substituting the result back into (22), we obtain

$$A^{N-M}_N = \frac{\epsilon^{N-M}}{M!}\widetilde{c}_2\Gamma(N-\beta)a^N\sum_{L=0}^{N-M}\frac{(N-L)!}{L!(N-L-M)!}$$

$$\times\left(-\frac{W_2}{a\epsilon}\right)^L J(N-L)(N-L)^{b+\beta}e^{S(L)}, \qquad (38)$$

where

$$J(N)=\int_0^\infty d\ln R^2 R^{-2}\exp\left(-Nf(R)+N\epsilon\ln R\right.$$

$$\left.+2K_d I_4(R)\frac{1-R^{-\epsilon}}{\epsilon}\right), \qquad (39)$$

$$S(L)=\frac{W_3}{\epsilon a^2 L}\left(\frac{N-L-M}{N-L}\right)^2$$

$$\times\left[1+\frac{(N-L)(N-M)}{L(N-M-L)}\ln\frac{N-L}{N}\right]. \qquad (40)$$

For $N - M \ll N$ or $N\epsilon \gg 1$, the sum over $L$ in (38) is determined by values $L \ll N$, and (38) goes over to (19). For $M \sim 1$, (38) becomes

$$A^{N-M}_N(\epsilon)=\frac{1}{M!}\epsilon^{N-M}\widetilde{c}_2\Gamma(N-\beta)a^N\sqrt{t/2\pi}\exp\left[f_\infty(Nt\ln N-1)+\frac{1}{t}\right]$$

$$\times\int_0^\infty dx\exp\left[-\frac{t}{2}\left(N-\frac{1}{t}-x\right)^2\right]$$

$$\times x^{M+b+\beta-f_\infty Nt}J(x), \qquad (41)$$

where

$$t=-\frac{\epsilon a}{W_2}\xrightarrow{\epsilon\to 0}\frac{3\epsilon}{n+8}, \quad f_\infty=\frac{W_3}{aW_2}\xrightarrow{\epsilon\to 0}\frac{3n+14}{n+8}. \qquad (42)$$

The assumptions made in the derivation of (41) are fulfilled in the region $Nt > 1$ or $1 - Nt \ll \epsilon^{1/2}$.

For $N\epsilon \ll 1$, the sum over $L$ in (34) is determined by the neighborhood of the upper limit of the sum, so that $\tau' \ll 1$; for $M \gg \ln N$ and $M \sim \ln N$, Eqs. (35) and (36) apply, respectively. For $M \ll \ln N$, terms with $s \gtrsim M$, $L = N - 2s$ dominate, and by virtue of the equality $A^0_0 = 1$ we have the following result for region II:

$$A^{N-M}_N(\epsilon)=(-W_2)^N\frac{\Gamma(N-\beta)}{\Gamma(N+1)\Gamma(-\beta)}\sum_{L=0}^\infty \epsilon^L\frac{1}{M!L!}$$

$$\times \left(-\frac{W_3}{W_2^2}\right)^{M+L} (N \ln N)^{M+L}, \qquad (43)$$

which can be obtained from the system of equations (11) by separating out the leading asymptotic behavior in $N$ for $A_N^{N-K,L}$. For $N\epsilon \ln N \ll 1$, terms with $s \leq M$ dominate, and for arbitrary $M$ we have the result

$$X_{N,M} = \frac{\Gamma(M-\beta)}{A_M^0} \sum_{s=0}^{M} \frac{A_{M-s}^0}{\Gamma(M-s-\beta)} \frac{1}{s!}$$

$$\times \left(\frac{W_3}{W_2} \ln \frac{N}{M-s+1}\right)^s e^{t(N-M)M}, \qquad (44)$$

whose region of applicability expands without limit as $\epsilon \to 0$, and this result transforms to Eqs. (42) and (43) of Ref. 16 for $d=4$.

## 4. ENERGY RENORMALIZATION AND DECAY

As in the case $d=4$ (Ref. 16), there are two important contributions to the sum (4)—a nonperturbative and a quasi-parquet contribution, arising respectively from regions I and II (Fig. 1). We restrict the discussion to the continuum limit $\Lambda \to \infty$, in which only the coefficients $A_N^N(\epsilon)$ remain in the sum (4). The quasiparquet contribution is calculated on the basis of formula (43), and has the form

$$[Y(\kappa)]_{\text{quasiparq}} = \left[\Delta + \frac{W_3(\epsilon)}{W_2(\epsilon)} u \kappa^{-\epsilon} \ln \Delta\right]^{\beta(\epsilon)},$$

$$\Delta \equiv 1 + W_2(\epsilon) u \frac{\kappa^{-\epsilon}}{\epsilon}, \qquad (45)$$

where the coefficients $W_2(\epsilon)$, $W_3(\epsilon)$, and $\beta(\epsilon)$ can be taken to zeroth order in $\epsilon$. Within the limits of accuracy of the calculations, the argument $\Delta$ of the logarithm can be replaced by its minimum value $\widetilde{\Delta} \sim \epsilon \ln \epsilon$ (defined by Eqs. (51) and (52) below), since for $\Delta \gg \widetilde{\Delta}$ the logarithmic term is unimportant. Therefore (45) can be rewritten in the form

$$[Y(\kappa)]_{\text{quasiparq}} = [1 + W_{2,0} \widetilde{u} \kappa^{-\epsilon}/\epsilon]^{\beta_0},$$

$$\widetilde{u} \equiv u \left[1 + \frac{W_{3,0}}{W_{2,0}^2} \epsilon \ln \widetilde{\Delta}\right], \qquad (46)$$

which differs from the parquet form[23] only by the substitution of $\widetilde{u}$ for $u$.

To calculate the nonperturbative contribution, we set

$$A_N^N(\epsilon) = \widetilde{c}_2 \Gamma(N+b) \epsilon^N a^N F(N) \qquad (47)$$

and sum (4) from some large $N_0$ to infinity according to Eq. (46) in Ref. 16:

$$[\Sigma(0,\kappa)]_{\text{nonpert}} \equiv i\Gamma_0(\kappa^2)$$

$$= i\pi \widetilde{c}_2 \kappa^2 (\kappa^\epsilon/au)^b e^{-\kappa^\epsilon/au} F(\kappa^\epsilon/au). \qquad (48)$$

The nonperturbative contribution is associated with the divergence of the series, and formally arises from the region of arbitrarily large $N$. However, it must be calculated on the basis of (41), not (19), since the correction factor distinguishing (41) from (19) is evaluated in effect for $N=\kappa^\epsilon/au$ and turns out to be substantial. I did not recognize this circumstance in Ref. 17; therefore, Eqs. (22) and (23) in Ref. 17 differ from Eqs. (52), (53), and (55) below.

Approximating the series (4) by the sum of contributions (46) and (48), we obtain

$$\kappa_0^2 - \kappa_c^2 = \kappa^2 [1 + 8K_4 \widetilde{u} \kappa^{-\epsilon}/\epsilon]^{1/4} + i\Gamma_0(\kappa^2),$$

$$\kappa^2 = -E - i\Gamma, \qquad (49)$$

where $\kappa_c^2 = \Sigma(0,0)$ and we have allowed for the fact that $\kappa_0^2 = \kappa^2 + \Sigma(0,\kappa)$. Equation (49) is solved like Eq. (93) in Ref. 15. Setting

$$\kappa^2 = |\kappa|^2 e^{-i\varphi}, \quad x = \frac{2}{\epsilon}\left[\left(\frac{|\kappa|^2}{\Gamma_c}\right)^{\epsilon/2} - 1\right], \quad \Gamma_c = \left(\frac{8K_4|\widetilde{u}|}{\epsilon}\right)^{2/\epsilon}$$
$$(50)$$

and separating the real and imaginary parts of (49), we obtain a connection between the decay $\Gamma$ and the renormalized energy $E$ with the unrenormalized energy $E_B = -\kappa_0^2$ in parametric form:

$$\Gamma = \Gamma_c \left(1 + \frac{\epsilon x}{2}\right)^{2/\epsilon} \sin \varphi, \quad E = -\Gamma_c \left(1 + \frac{\epsilon x}{2}\right)^{2/\epsilon} \cos \varphi,$$

$$-E_B + E_c = \Gamma_c \left(1 + \frac{\epsilon x}{2}\right)^{2/\epsilon} \left(\frac{\epsilon x/2}{1+\epsilon x/2}\right)^{1/4} \left[\cos\left(\varphi + \frac{\varphi}{4x}\right)\right.$$

$$\left. - \tan \frac{\varphi(1+2\epsilon x)}{3} \sin\left(\varphi + \frac{\varphi}{4x}\right)\right], \qquad (51)$$

where $E_c$ is defined by Eq. (108) in Ref. 15, and $x(\varphi)$ is a single-valued function in the interval $0 < \varphi < \pi$, analogous to the function shown in Fig. 2 of Ref. 15, and implicitly defined by the equation

$$\sin\left(\varphi + \frac{\varphi}{4x}\right) = \frac{e^{-4x/3}}{x^{1/4}} I(x) \cos \frac{\varphi(1+2\epsilon x)}{3}, \qquad (52)$$

where

$$I(x) = \widetilde{c}_2 \left(\frac{3}{4}\right)^{1/4} \left(\frac{\pi t}{2}\right)^{1/2}$$

$$\times \exp\left\{-f_\infty + f_\infty \left(1 + \frac{\epsilon x}{2}\right) \ln\left[\widetilde{\Delta}\left(1 + \frac{\epsilon x}{2}\right) \bigg/ t\right]\right\}$$

$$\times \int_0^\infty dz \exp\left[-\frac{t}{2}\left(\frac{\epsilon x}{2t} - z\right)^2\right] z^{b+\beta-f_\infty(1+\epsilon x/2)} J(z).$$

$$(53)$$

Equations (51) and (52) simplify substantially in two overlapping regions. For $x \gg \ln(1/\epsilon)$, i.e., at high $|E|$, where the right-hand side of Eq. (52) is small and the quantity $\varphi$ is near 0 or $\pi$, we obtain the asymptotic behavior of $\Gamma(E)$,

$$\Gamma(E) = \begin{cases} \dfrac{1}{8} \pi \epsilon E[(E/\Gamma_c)^{\epsilon/2} - 1]^{-1}, & E \gg \Gamma, \\ \Gamma_0(E)[1 - (|E|/\Gamma_c)^{-\epsilon/2}]^{-1/4}, & -E \gg \Gamma, \end{cases} \qquad (54)$$

which produce the illusion of a ghost pole[7] ($\Gamma_0(E) \equiv \Gamma_0(|\kappa|^2)$). For large positive $E$ the result of the kinetic equation is reproduced; for large negative $E$ the decay becomes purely nonperturbative.

At low energies, $x \lesssim \epsilon^{-1/2}$, we have

$$\sin\left(\varphi + \frac{\varphi}{4x}\right) = I(0) \frac{e^{-4x/3}}{x^{1/4}} \cos\frac{\varphi}{3},$$

$$I(0) \sim \epsilon^{-7/12} \left(\ln\frac{1}{\epsilon}\right)^{17/12}, \quad (55)$$

which describes the neighborhood of the ghost pole and has the same functional form as the four-dimensional equation (see Eq. (51) in Ref. 16 for $x \ll x_0$ and Eq. (100) in Ref. 15). The minimum values of $\Delta$ and $x$ are reached simultaneously, and to logarithmic accuracy they are

$$\Delta_{\min} \equiv \widetilde{\Delta} \approx \frac{7}{8} \epsilon \ln\frac{1}{\epsilon}, \quad x_{\min} \approx \frac{7}{16} \ln\frac{1}{\epsilon}, \quad (56)$$

so that the detour about the pole must be taken at a distance of the order of $\epsilon \ln(1/\epsilon)$.

## 5. DENSITY OF STATES

To calculate the density of states requires a knowledge of the eigenenergy $\Sigma(p,\kappa)$ for finite momenta[15]; like $p=0$, this quantity consists of a nonperturbative and a quasiparquet contribution. The quasiparquet contribution is given by the parquet equations (Ref. 15, Sec. 7) with the substitution $u \to \hat{u}$; the proof of this is completely analogous to the situation $d=4$ (Ref. 16, Sec. 5). The nonperturbative contribution turns out to be important only in the region of large negative $E$, where it is directly determined by the Lipatov asymptotic behavior, and can be calculated on the basis of formula (13) (for $N = \kappa^\epsilon/au \gg 1/\epsilon$ the correction factor distinguishing results of the type (41) and (19) is equal to unity)

$$[\Sigma(p,\kappa)]_{\text{nonpert}} = i\pi c_2 \kappa^2 \left(\frac{\kappa^\epsilon}{au}\right)^b e^{-\kappa^\epsilon/au}$$

$$\times \int_0^\infty d\ln R^2 R^{-2} \langle \phi_c^3 \rangle_{pR/\kappa} \langle \phi_c^3 \rangle_{-pR/\kappa}$$

$$\times \exp\left\{-\frac{\kappa^\epsilon}{au}[f(R) - \epsilon \ln R]\right.$$

$$\left. + \frac{2K_d I_4(R)}{\epsilon}\right\}. \quad (57)$$

For $p=0$ the integral is governed by the neighborhood of the saddle point $R_0$, which is a root of the equation

$$\epsilon = 6R_0^2(-\ln R_0 + \ln 2 - C - 1), \quad (58)$$

so that $R_0 \approx \sqrt{\epsilon/3 \ln(1/\epsilon)}$. For $p \lesssim \kappa R_0^{-1}$, Eq. (57) does not depend on $p$; for $p \gtrsim \kappa R_0^{-1}$ it falls off rapidly with increasing $p$. By virtue of the logarithmic accuracy of the following calculations (Ref. 15, Sec. 8) the result

$$[\Sigma(p,\kappa)]_{\text{nonpert}} \approx [\Sigma(0,\kappa)]_{\text{nonpert}} \theta(\kappa R_0^{-1} - p). \quad (59)$$

suffices. Taking the above into account, the final expression for $\Sigma(p,\kappa)$ has the form

$$\Sigma(p,\kappa) - \Sigma(0,\kappa) = \kappa^2 \left\{1 - \frac{3}{2}\left[\frac{t(x)}{t(x_\infty)}\right]^{-1/4} + \frac{1}{2}\left[\frac{t(x)}{t(x_\infty)}\right]^{-3/4}\right\}$$

$$- i\Gamma_0(\kappa^2) \theta(p - \kappa R_0^{-1}) \quad (60)$$

(cf. Eq. (116) in Ref. 15), where

$$t(x) = 1 + 8K_4 \widetilde{u} x/\epsilon, \quad x = p^{-\epsilon}, \quad x_\infty = \kappa^{-\epsilon}. \quad (61)$$

Substituting (60) into Eqs. (117) and (118) of Ref. 15 for $d = 4 - \epsilon$, we obtain

$$\nu = \frac{\Gamma_c}{4\pi|\widetilde{u}|}\left(1 + \frac{\epsilon x}{2}\right)^{2/\epsilon}\left\{\left(1 + \frac{2}{\epsilon x}\right)^{-1/4}\left(1 - \frac{R_0^\epsilon}{2 + \epsilon x}\right)\right.$$

$$\left. \times \sin\left(\varphi + \frac{\varphi}{4x}\right) - \left(1 + \frac{2}{\epsilon x}\right)^{-3/4} \sin\left(\varphi + \frac{3\varphi}{4x}\right)\right\}, \quad (62)$$

which together with (51) and (52), determines the density of states $\nu(E)$ in parametric form.

Let us now turn our attention to the presence of scaling: for the energy measured in units of $\Gamma_c$ and the density of states in units of $\Gamma_c/|\widetilde{u}|$, all dependences are determined by universal functions that are independent of the degree of disorder. For $|E| \gg \Gamma$, we have the asymptotic behavior

$$\nu(E) = \begin{cases} \dfrac{1}{2} K_4 E^{(d-2)/2}\left[1 - \left(\dfrac{E}{\Gamma_c}\right)^{-\epsilon/2}\right]^{-1/4}, & E \gg \Gamma, \\ \dfrac{\Gamma_0(E)}{4\pi|\widetilde{u}|}\left\{1 - \dfrac{R_0^\epsilon}{2}\left(\dfrac{|E|}{\Gamma_c}\right)^{-\epsilon/2} - \left[1 - \left(\dfrac{|E|}{\Gamma_c}\right)^{-\epsilon/2}\right]^{1/2}\right\}, \\ -E \gg \Gamma, \end{cases} \quad (63)$$

indicating a ghost pole. For large positive $E$, the function $\nu(E)$ transforms into the density of states of an ideal system, and at large negative energies $E$ we obtain the following result for the fluctuation tail:

$$\nu(E) = \frac{K_4}{\pi} \Gamma_0(E) |E|^{-\epsilon/2} \ln\frac{1}{R_0}$$

$$= \widetilde{c}_2 K_4 \left(\frac{2\pi}{3} \ln\frac{1}{R_0}\right)^{1/2} R_0^{-3} |E|^{(d-2)/2}\left[\frac{\widetilde{I}_4 |E|^{\epsilon/2}}{4|u|}\right]^{(d+1)/2}$$

$$\times \exp\left(\frac{2K_d I_4(R_0)}{\epsilon} - \frac{I_4(R_0)|E|^{\epsilon/2}}{4|u|R_0^\epsilon}\right), \quad (64)$$

whose energy dependence coincides with that obtained in Refs. 25–27, and corresponds to the well-known Lifshits law[28]; the discrepancy at $\epsilon \to 0$ is eliminated for finite cutoff parameter $\Lambda$. Oddly enough, for $\epsilon x \ll 1$ Eqs. (51)–(53) and (62) have the same functional form as those for $d = 4$ (Ref. 16), i.e., the behavior of all physical quantities in the vicinity of the mobility threshold turns out to be effectively four-dimensional. As in Refs. 15 and 16, the phase transition point is shifted into the complex plane, which ensures regularity of the density of states at all energies.

$R_0^\epsilon$ differs substantially from unity only when $\kappa^\epsilon/u \ll 1/\epsilon$, the terms in braces in (63) cancelling almost exactly. Letting $R_0 \to 1$ in (60) is tantamount to completely neglecting $[\Sigma(p,\kappa)]_{\text{nonpert}}$, since the domain of integration in Eq. (118) of Ref. 15 is $p \gtrsim \kappa$. Thus, $[\Sigma(p,\kappa)]_{\text{nonpert}}$ is significant only for large-magnitude negative $E$, and can be calculated using the Lipatov asymptotic form.

## 6. LIPATOV ASYMPTOTIC LIMIT

Calculation of the Lipatov asymptotic limit in $(4-\epsilon)$-theory closely follows the scheme for $d=4$ described in detail in Ref. 16. Therefore we discuss only the differences that arise, denoting by the numeral I references to equations from Ref. 16.

In massless four-dimensional theory there exists a specific zero mode—the dilatation mode, corresponding to variation of the radius $R$ of the instanton.[16,18,29] As in the massive four-dimensional theory,[16] for $d=4-\epsilon$ this mode becomes soft and the integration over it bears a substantially non-Gaussian character. It is necessary to carry out this integration correctly to ensure that the correct limit is reached as $d \to 4$.

By analogy with (I.82), we introduce three expansions of unity inside the functional integral:

$$1 = \left(\int d^d x |\varphi(x)|^4\right)^d \int d^d x_0$$
$$\times \prod_{\mu=1}^{d} \delta\left(-\int d^d x |\varphi(x)|^4 (x-x_0)_\mu\right),$$

$$1 = \int d^d x |\varphi(x)|^4 \int_0^\infty d\ln R^2$$
$$\times \delta\left(-\int d^d x |\varphi(x)|^4 \ln\left(\frac{x-x_0}{R}\right)^2\right),$$

$$1 = \int d^n u\, \delta(\mathbf{u} - \mathbf{v}\{\varphi\}), \tag{65}$$

and in place of (I.82) we make the substitutions

$$x - x_0 = R\tilde{x}, \quad \varphi_\alpha(x_0 + R\tilde{x}) = R^{-(d-2)/2} \tilde{\varphi}_\alpha(\tilde{x}),$$
$$g = \tilde{g} R^{d-4}. \tag{66}$$

As a result, we have

$$[G_M]_{N-1} = \int_0^\infty d\ln R^2 Z_0(\kappa_R)^{-1}$$
$$\times \int d^d x_0 \int d^n u R^{-4-(d-2)M/2} \int \frac{dg}{2\pi i}$$
$$\times \int D\varphi \prod_{\mu=1}^{d} \delta\left(-\int d^d x |\varphi(x)|^4 x_\mu\right)$$
$$\times \delta\left(-\int d^d x |\varphi(x)|^4 \ln x^2\right) \delta(\mathbf{u}-\mathbf{v})$$
$$\times \left(\int d^d x |\varphi(x)|^4\right)^{d+1} \varphi_{\alpha_1}\left(\frac{x_1-x_0}{R}\right) \ldots \varphi_{\alpha_M}$$
$$\times \left(\frac{x_M - x_0}{R}\right) \exp[-H\{\kappa_R, g, \varphi\} - N\ln g$$
$$+ N\epsilon \ln R], \tag{67}$$

the main difference of which from (I.83) consists in the appearance of the term $N\epsilon \ln R$ in the exponential. The choice of instanton, as before, is dictated by Eq. (I.94), which after transforming to the function $\phi_c(x)$ according to (I.72) takes the following form in spherical coordinates ($r \equiv |x|$):

$$\phi_c''(r) + \frac{3-\epsilon}{r} \phi_c'(r) - \kappa_R^2 \phi_c(r) + \phi_c^3(x) - \mu_0 \phi_c^3(r) \ln r^2 = 0. \tag{68}$$

In the region $r \ll \kappa_R^{-1}$, terms with $\epsilon$, $\kappa_R$, and $\mu_0$ are treated as a perturbation, and by analogy with (I.99) we obtain

$$\phi_c(r) = \frac{2\sqrt{2}}{z+1} \left[1 + \frac{1-z}{1+z} v(z)\right]_{z=r^2},$$

$$v(z) = \int_0^z dz \frac{(1+z)^4}{(1-z)^2 z^2} \left\{ \epsilon \frac{z^2(z-3)}{12(1+z)^3} \right.$$
$$+ \frac{\kappa_R^2}{4} \left[-\ln(1+z) + \frac{z+2z^2}{(1+z)^2}\right]$$
$$+ \left. \mu_0 \left[\frac{\ln z}{(z+1)^4} - \frac{z+3}{6(z+1)^3}\right] z^2 \right\}. \tag{69}$$

Calculation of the asymptotic limit of $v(z)$ for $z \gg 1$ with allowance for only the growing terms in $z$ gives for the region $1 \ll r \ll \kappa_R^{-1}$

$$\phi_c(r) = \frac{2\sqrt{2}}{r^2} \left\{1 + \frac{1}{2} \kappa_R^2 r^2 \ln r + \left[\frac{1}{6} \mu_0 - \frac{3}{4} \kappa_R^2 - \frac{1}{12} \epsilon\right] r^2 \right.$$
$$\left. + 3\kappa_R^2 \ln^2 r + \left[2\mu_0 - \frac{11}{2} \kappa_R^2\right] \ln r - \frac{1}{r^2}\right\}. \tag{70}$$

In the region $r \gg 1$, treating the nonlinear terms in (68) as a perturbation, we obtain after separating out the asymptotic limit for $r \ll \kappa_R^{-1}$

$$\phi_c(r) = \frac{2\sqrt{2}}{r^2} \left\{1 + \frac{1}{2} \kappa_R^2 r^2 \ln r \right.$$
$$+ \frac{2C - 1 + 2\ln(\kappa_R/2)}{4} \kappa_R^2 r^2 + 3\kappa_R^2 \ln^2 r$$
$$\left. + \left[\epsilon + \kappa_R^2\left(6C + \frac{1}{2} + 6\ln\frac{\kappa_R}{2}\right)\right] \ln r - \frac{1}{r^2}\right\}. \tag{71}$$

The matching condition for (70) and (71) has the form

$$2\mu_0 = \epsilon + 6\kappa_R^2 (\ln \kappa_R + C + 1 - \ln 2) \tag{72}$$

Using Eq. (69) to calculate the integral in (I.70) (making the substitution $d^4 x \to d^d x$), we obtain

$$N \ln g_c = N \ln\left(-\frac{\bar{I}_4}{4N}\right) + N f(\kappa_R), \tag{73}$$

where $f(x)$ and $\bar{I}_4$ are defined by (14). In comparison with the case $d=4$, the function $f(x)$ differs by a constant $\sim \epsilon$.

Another modification arises when the divergences are separated out of the determinants defined by the sum rule [cf. (I.114)]

$$\sum_s \frac{1}{\mu_s^2} = 9 \int_0^{\Lambda R} \frac{d^d k}{(2\pi)^d}$$

$$\times \int_0^{\Lambda R} \frac{d^d q}{(2\pi)^d} \frac{\langle \phi_c^2 \rangle_q \langle \phi_c^2 \rangle_{-q}}{(k^2 + \kappa_R^2)[(k+q)^2 + \kappa_R^2]}$$

$$\approx 9 K_d I_4(\kappa_R) \frac{1 - (\Lambda R)^{-\epsilon}}{\epsilon} + 12\left(\frac{1}{3} + C - \ln 2\right). \quad (74)$$

For the $N$th coefficient of the expansion of the Green's function, instead of (I.113) we obtain

$$[G_M(x_1, \alpha_1, \ldots, x_M, \alpha_M)]_N$$

$$= c(-1)^N \left(\frac{4}{\overline{I}_4}\right)^{N+(M+d+1)/2} \Gamma\left(N + \frac{M+n+d}{2}\right)$$

$$\times \int d^n u \, \delta(|\mathbf{u}| - 1) u_{\alpha_1} \ldots u_{\alpha_M} \int_0^\infty d \ln R^2$$

$$\times \int d^d x_0 R^{-d - M(d-2)/2} \phi_c\left(\frac{x_1 - x_0}{R}\right) \ldots \phi_c\left(\frac{x_M - x_0}{R}\right)$$

$$\times \exp\left\{-N f(\kappa_R) + N \epsilon \ln R\right.$$

$$+ \frac{n+8}{4} K_d I_4(\kappa_R) \frac{1 - (\Lambda R)^{-\epsilon}}{\epsilon}\right\}, \quad (75)$$

where the constant $c$ is calculated in the lowest order in $\epsilon$ and is given by formula (I.114). Going over to the vertex part, instead of (I.127) we obtain

$$[\Gamma^{(0,2M)}(p_1, \ldots, p_{2M})]_N$$

$$= c(-1)^N \frac{2\pi^{n/2}}{2^M \Gamma(M+n/2)} \left(\frac{4}{\overline{I}_4}\right)^{N+M+5/2}$$

$$\times \Gamma\left(N + \frac{2M+n+d}{2}\right) \int_0^\infty d \ln R^2 R^{-d + (d-2)M}$$

$$\times \langle \phi_c^3 \rangle_{Rp_1} \ldots \langle \phi_c^3 \rangle_{Rp_{2M}} \exp\left\{-N f(\kappa_R) + N \epsilon \ln R\right.$$

$$+ \frac{n+8}{4} K_d I_4(\kappa_R) \frac{1 - (\Lambda R)^{-\epsilon}}{\epsilon}\right\}, \quad (76)$$

where $\langle \phi_c^3 \rangle_p$ is the Fourier component of the function $\phi_c^3(x)$. To lowest order in $\epsilon$ this Fourier component is given by (14). The vertex $\Gamma^{(0,2)}$ coincides with the eigenenergy and for $M=1$, $n=0$, (76) follows from (13).

## 7. INSTANTON RESULTS FOR $\epsilon \sim 1$

In order to compare with the results of other authors,[7,25–27] let us discuss instanton calculations for $d < 4$ without assuming that $d$ is close to 4. Such calculations closely follow the scheme for $d > 4$ described in Ref. 15 (we denote references to the corresponding formulas by the numeral II), with the replacements $\Sigma_x \to \int d^d x$ and $\epsilon(p) \to p^2$. The difference has to do with the need to separate out the zero translational modes along with the rotational modes; the dilatation mode is considered here, in contrast to the previous section, on general grounds. Accordingly, of the three expansions of unity (65) we use only the first and the third, but the substitution of variables (66) is carried out with $R = 1$. In addition to (II.65), a transformation of the determinant $D_L$ is required:

$$\frac{D_L'}{D_0} = \bar{D}(1) \prod_{\mu=1}^d \frac{\int d^d x \left(\frac{\partial \phi_c(x)}{\partial x_\mu}\right)^2}{3 \int d^d x \, \phi_c^2(x) \left(\frac{\partial \phi_c(x)}{\partial x_\mu}\right)^2},$$

$$\bar{D}(1) = \prod_s{}' \left(1 - \frac{1}{\mu_s}\right). \quad (77)$$

The prime denotes omission of the contribution of the translational modes. The instanton equation reduces by this substitution of variables to the form

$$\Delta \phi_c(x) + \phi_c^3(x) - \bar{\kappa}^2 \phi_c(x) = 0, \quad (78)$$

where $\bar{\kappa}$ is an arbitrary parameter (see below). For the expansion coefficients of the Green's function we obtain

$$[G_M(x_1, \alpha_1, \ldots, x_M, \alpha_M)]_N$$

$$= \frac{2^{n-1}}{(2\pi)^{(n+d+1)/2}} \left(\frac{I_6 - \bar{\kappa}^2 I_4}{d}\right)^{d/2} \left(\frac{4}{I_4}\right)^{(M+d)/2}$$

$$\times \left(\frac{\kappa}{\bar{\kappa}}\right)^{(d-2)M/2} \left[-\bar{D}(1) \bar{D}^{n-1}(1/3)\right]^{-1/2}$$

$$\times \left[-\frac{4}{I_4} \left(\frac{\kappa}{\bar{\kappa}}\right)^{d-4}\right]^N \Gamma\left(N + \frac{M+n+d-1}{2}\right)$$

$$\times \int d^d x_0 \phi_c\left(\frac{\kappa}{\bar{\kappa}} x_1 - x_0\right) \ldots \phi_c\left(\frac{\kappa}{\bar{\kappa}} x_M - x_0\right)$$

$$\times \int d^n u \, \delta(|\mathbf{u}| - 1) u_{\alpha_1} \ldots u_{\alpha_M}, \quad (79)$$

where

$$I_p = \int d^d x \, \phi_c^p(x). \quad (80)$$

For $d = 4 - \epsilon$, (75) and (79) are equivalent only for $N\epsilon \gg 1$, when the integration over $R$ in (75), corresponding to the dilatation mode, can be carried out in the saddle-point approximation. The saddle point occurs for $\kappa_R = R_0$, where $R_0$ is a root of Eq. (58). In this case, by virtue of Eq. (72), we have $\mu_0 = 0$ and the instanton equation (68) reduces to Eq. (78) with $\bar{\kappa} = R_0$. Expression (79) with $\bar{\kappa} = R_0$, after estimating the pre-exponential in the zeroth order in $\epsilon$, differs from the result for the saddle-point approximation in (75) by the constant factor

$$\left[\frac{\lambda_0^L + R_0^2}{3 I_4 R_0^2(-\ln R_0 + \ln 2 - C - 3/2)} \int d^d x [e_0^L(x)]^2\right]^{1/2}$$

$$\approx \left[ \frac{R_0^2 - \rho^2}{R_0^2} \frac{\ln \rho}{\ln \epsilon} \right]^{1/2}, \tag{81}$$

where $\lambda_0^L \equiv -\rho^2$ and $e_0^L(x)$ are the eigenvalue and eigenfunction of the operator $-\Delta - 3\phi_c^2(x)$ corresponding to the dilatation mode. The normalization of the function $e_0^L(x)$ is chosen so as to coincide with $[\partial \phi_c(x)/\partial R]_{R=1}$ in the region $|x| \lesssim 1$. The quantity (81) is equal to unity for $\rho \sim \epsilon$ or $-\lambda_0^L \sim \epsilon^2$; from perturbation theory it is easy to convince oneself that the contribution to $\lambda_0^L$ that is first-order in $\epsilon$ vanishes, due to the divergence of the normalization integral for $e_0^L(x)$ for $d=4$.

For $2 \leq d < 4$, the determinants $\bar{D}(1)$ and $\bar{D}(1/3)$ contain divergences,[25] which can be eliminated by renormalization according to (II.75) with simultaneous transformation to the renormalized energy $E$ (the Thomas–Fermi method yields $\mu_s \sim s^{2/d}$ for $s \gg 1$, and the first sum in (II.69) diverges). Setting $\bar{\kappa} = 1$ and summing the non-leading terms of the perturbation-theory series for the two-point ($M=2$) Green's function according to (II.90), it is not hard to obtain an expression for the fluctuation tail of the density of states:

$$\nu(E) = \frac{(4-d)2^{d-1}}{(2\pi)^{(d+1)/2}} \left( \frac{I_6 - I_4}{I_4 d} \right)^{d/2} \left| \frac{\bar{D}_R(1/3)}{\bar{D}_R(1)} \right|^{1/2} |E|^{(d-2)/2}$$
$$\times \left( \frac{I_4 |E|^{(4-d)/2}}{2 a_0^d W^2} \right)^{(d+1)/2} \exp\left( -\frac{I_4 |E|^{(4-d)/2}}{2 a_0^d W^2} \right) \tag{82}$$

(where $4I_2 = (4-d)I_4$). The energy dependence of this expression for the fluctuation tail of the density of states coincides with that obtained by Cardy.[27] Normalization to the unperturbed density of states $\nu_0(E)$ and changing over from the renormalized energy $E$ to the unrenormalized energy $E_B$ with a simultaneous shift of the origin (see formula (12) in Ref. 26) gives the results of Brézin and Parisi (Ref. 26)[1)]

$$\frac{\nu(E_B)}{\nu_0(-E_B)} = \left( \frac{I_6 - I_4}{3} \right)^{3/2} \left| I_4 \frac{\bar{D}_R(1/3)}{\bar{D}_R(1)} \right|^{1/2} \frac{|E_B|}{(a_0^d W^2)^2}$$
$$\times \exp\left( -\frac{I_4}{16\pi} - \frac{I_4 |E_B|^{1/2}}{2 a_0^d W^2} \right), \quad d=3,$$

$$\frac{\nu(E_B)}{\nu_0(-E_B)} = \frac{I_6 - I_4}{8\pi^2} \left| I_4 \frac{\bar{D}_R(1/3)}{\bar{D}_R(1)} \right|^{1/2} \left( \frac{4\pi |E_B|}{a_0^d W^2} \right)^{3/2 - I_4/8\pi}$$
$$\times \exp\left( -\frac{I_4}{8\pi} - \frac{I_4 |E_B|}{2 a_0^d W^2} \right), \quad d=2. \tag{83}$$

For $d < 2$, there are no divergences in the determinants, and (82) holds in terms of the unrenormalized quantities (i.e., after the substitutions $E \to E_B$, $\bar{D}_R(1) \to \bar{D}(1)$, and $\bar{D}_R(1/3) \to \bar{D}(1/3)$). For $d=1$, Eq. (78) with $\tilde{\kappa} = 1$ has the solution $\phi_c(x) = \sqrt{2}/\cosh x$, and Eq. (II.64)

$$y'' - y + \frac{\mu_s}{\cosh^2 x} y = 0 \tag{84}$$

has eigenvalues $\mu_s = s(s+1)$, $s=1,2,\ldots$ since by the substitution $y = \tilde{y} \cosh^{-s} x$ it reduces to a form analogous to (I.121). Calculation of the parameters entering into (82)

$$\bar{D}(1) = \prod_{\substack{s=1 \\ s \neq 2}}^{\infty} \frac{(s+3)(s-2)}{s(s+1)} = -\frac{1}{5},$$

$$\bar{D}(1/3) = \prod_{s=2}^{\infty} \frac{(s+2)(s-1)}{s(s+1)} = \frac{1}{3},$$

$$I_4 = \frac{16}{3}, \quad I_6 = \frac{128}{15} \tag{85}$$

yields the result

$$\nu(E_B) = \frac{4}{\pi} \frac{|E_B|}{a_0^d W^2} \exp\left\{ -\frac{8|E_B|^{3/2}}{3 a_0^d W^2} \right\}, \tag{86}$$

which agrees with the exact solution due to Halperin.[10,30]

### ACKNOWLEDGMENTS

I am grateful to the participants of the seminars at the Institute of Semiconductor Physics (IFP) and the Physics Institute of the Academy of Sciences (FIAN) for their interest in this work.

This work was carried out with the financial support of the International Science Foundation and the Russian Government (Grants No. MON 000 and No. MON 300) and the Russian Fund for Fundamental Research (Project No. 96-02-19527).

---

[1)]The left-hand sides of the final formulas (16) of Ref. 26 contain obvious typographical errors; substitution in expression (83) of the numerical values of the parameters obtained in Ref. 25 yields the coefficients shown in Ref. 26 on the right-hand sides of (16).

Translated by Paul F. Schippnick